%% file: sample-sigconf.tex
\documentclass[sigconf]{acmart}
\AtBeginDocument{%
  }

\usepackage{amsmath}
\usepackage{graphicx}
\usepackage{subcaption}

\newcommand{\cpp}{C\texttt{++} }

\usepackage{listings}
\usepackage{color}

\definecolor{dkgreen}{rgb}{0,0.6,0}
\definecolor{gray}{rgb}{0.5,0.5,0.5}
\definecolor{mauve}{rgb}{0.58,0,0.82}

\setcopyright{acmlicensed}
\copyrightyear{2025}
\acmYear{2025}
\acmDOI{}
\acmBooktitle{Workshop on Compilers, Deployment, and Tooling for AI (CODAI'25), January 20, 2025, Barcelona, Spain}
\acmISBN{}

\begin{document}

\title{A lightweight C Code Generator for Heterogeneous TinyML Systems}




\author{Chen Liu$^*$, Bernhard Vogginger$^*$, Christian Mayr$^{* \dagger}$}
\affiliation{%
  \department{$^*$Chair of Highly-Parallel VLSI-Systems and Neuro-Microelectronics}
  \department{$^\dagger$Centre for Tactile Internet with Human-in-the-Loop (CeTI)}
  \institution{Dresden University of Technology}
  \city{Dresden}
  \country{Germany}
}
\email{{chen.liu, bernhard.vogginger, christian.mayr}@tu-dresden.de}





\renewcommand{\shortauthors}{Liu et al.}

\begin{abstract}
Over the past decade, the significant advancements in artificial intelligence have been closely tied to increasingly sophisticated training frameworks and steadily powerful computing hardware.
However, the entire ecosystem remains fragmented.
On one hand, models are represented with frameworks dependent formats, while on the other hand, new tailored hardware for accelerating novel operators has continued to emerge.
To effectively diminish the transfer costs from training to inference, we propose a lightweight C code generator designed for bare-metal heterogeneous systems that incorporate domain-specific accelerators.
Through a comprehensive case study on a RISC-V based radar DSP, we demonstrate that by employing an accelerator with ahead-of-time optimizations like weight tuning and bias fusion, our approach achieves a $2.3 \times$ improvement in inference speed and a 20\% reduction in binary size compared to the microTVM flow registered with the same accelerator,
thus offering another compelling choice for resource-constrained edge devices.
The code is open-sourced in Gitlab \footnote{https://gitlab.com/chensterliu/dnn\_codegen}.
\end{abstract}



\maketitle

\input{main}
\begin{acks}
This work was funded by the German Federal Ministry of Education and Research (BMBF), funding reference 16ME0729K, joint project ”EVENTS”, and
by the German Research Foundation (DFG, Deutsche Forschungsgemeinschaft) as part of Germany’s Excellence Strategy – EXC 2050/1 – Project ID 390696704 – Cluster of Excellence “Centre for Tactile Internet with Human-in-the-Loop” (CeTI) of Technische Universität Dresden.
\end{acks}

\bibliographystyle{ACM-Reference-Format}
\bibliography{sample-base, main}










\end{document}

%% file: main.tex
\section{Introduction}


Numerous lightweight deep learning models have recently emerged, achieving impressive performance in classification and recognition tasks while consuming minimal resources. Deploying these models on edge devices and serving in low-power mode have become increasingly popular. Established software frameworks, such as TensorFlow \cite{tensorflow2015-whitepaper} and PyTorch \cite{paszke2019pytorch}, along with general-purpose hardware (e.g., CPUs, GPUs, and TPUs), support model training in a mature manner. However, when it comes to deploying deep learning models on a wide variety of edge devices with highly diverse hardware resources, the deployment process becomes complex, and efficiency cannot be guaranteed—this includes both model runtime performance and human resource expenditure.

To address these challenges, major frameworks have developed expended toolkits to simplify deployment. For instance, Google introduced TensorFlow Lite (TFLite) \cite{tflite} and TensorFlow Lite Micro (TFLM) \cite{david2021tensorflow} to enable production-level model deployment on resource-constrained devices. Similarly, Meta has announced ExecuTorch \cite{executorch} to enhence its deployment pipeline for PyTorch.

Another force to overcoming deployment barriers comes from device vendors, who provide end-to-end development toolkits optimized for their own hardware. These toolkits ensure model performance on deterministic backend devices, such as STM32Cube.AI from STMicroelectronics \cite{stm32cubeai}, ai8x-synthesis from Analog Devices \cite{ai8x}, and eIQ toolkit from NXP \cite{eiq}.

Multi-Level Intermediate Representation (MLIR) is an emerging framework for creating reusable and extensible compiler infrastructures. Its versatility has been demonstrated in diverse applications, including CPU-FPGA heterogeneous computing nodes \cite{10653659} and FPGA-based data centers 
\cite{pilato2021everest}. 
Building on MLIR’s capabilities, the Tiny Intermediate Representation Execution Environment (TinyIREE) framework further generates compact workloads with optimized runtime libraries for embedded systems \cite{9782563}. Additionally, frameworks like TVM \cite{chen2018tvm} have also been employed to deploy DNN models on heterogeneous edge devices, expanding the scope of efficient model deployment in edge computing \cite{hamdi2024match}.

However, current solutions still face several limitations:
\begin{itemize}
    \item  While TFLM and ExecuTorch remove OS dependencies, they continue to rely on \cpp to wrap low-level library functions. An embedded market study \cite{2019study} indicates that C usage still dominates embedded systems, being over twice as prevalent as \cpp. The binary size generated by \cpp remains prohibitively large for tiny devices.
    
    \item  Device vendors’ commercial tools are highly fragmented, as they are optimized exclusively for their proprietary hardware. This approach limits user intervention in the deployment process and creates a strong dependency on specific backend devices.

    \item Deployment tools offer limited support for domain-specific accelerators (DSAs), especially custom-built DSAs, restricting the flexibility required for specific applications.

    \item Integrating new operators into the MLIR ecosystem involves significant overhead, including building dialects, patterns, abstraction lowering, etc. 

    \item Model weights are quantized to lower precision integers, but the associated overhead of floating-point based rescaling operations is often overlooked.
\end{itemize}

To address above limitations, we propose a lightweight C code generator designed to overcome existing challenges. Our tool accepts various trained model formats, applies optimization techniques like weight tuning and bias fusion, provides C function templates for customized DSAs, and generates human-readable C source code to enhance compatibility and transparency. This approach facilitates efficient model deployment across diverse edge devices while supporting both standard and specialized hardware configurations.


\section{Implementation}

Our C code generator, accessible via a Python interface, is designed for ease of use and follows two primary design principles.
\paragraph{Pure C Code Generation} The generator produces pure C code compatible with the vast majority of embedded systems, ensuring high compatibility even for non-commercial, research-oriented devices. The streamlined compiled binary size is especially suitable for memory-constrained tiny devices and custom-designed domain-specific accelerators.

\paragraph{Template Mechanism for backend Integration} By employing a template-based approach, the generator seamlessly maps model operators to C functions for the backend, enabling an end-to-end model parsing pipeline. The resulting code is both human-readable and editable, allowing for extensive customization in experimental deployments. A key advantage of the template method is that generation stops at source code, leaving the final compilation to be handled by the backend hardware’s toolchain, thus minimizing potential compilation errors.

\subsection{Overview}

\begin{figure}[htbp]
\centering
\includegraphics[width=0.9\linewidth]{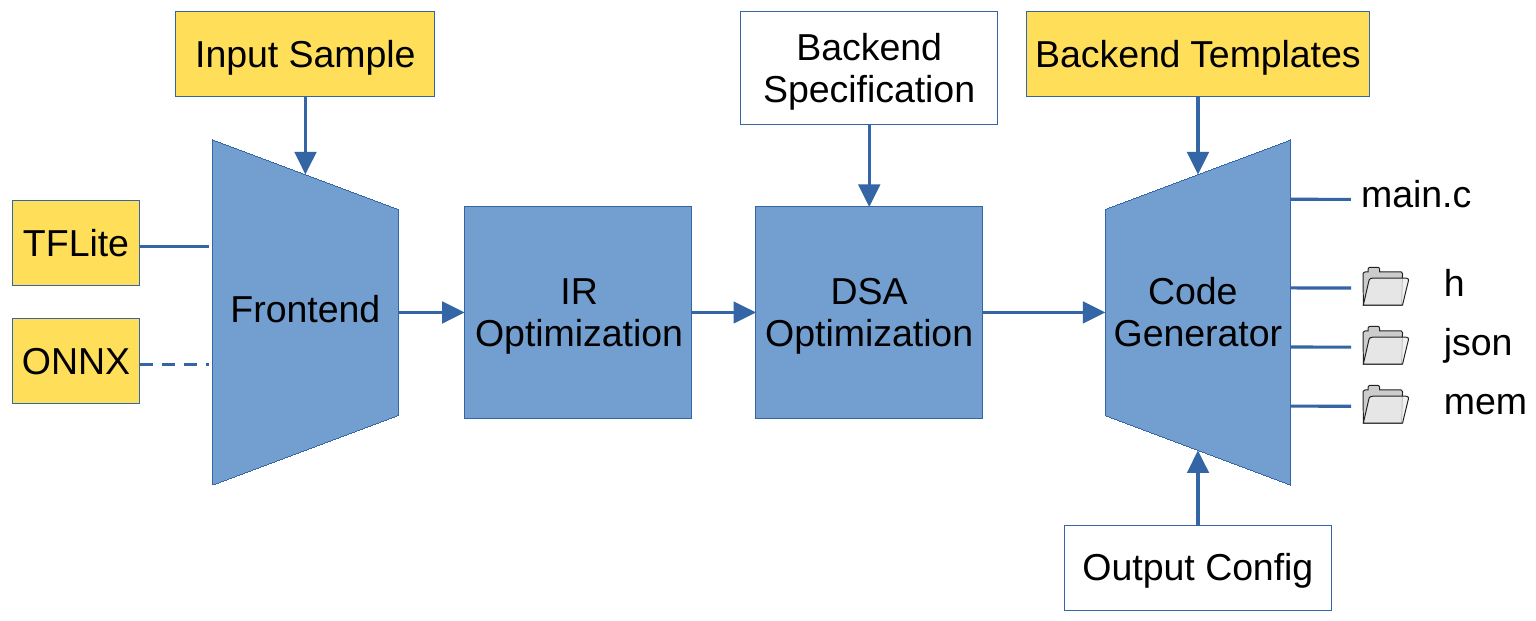}
\caption{\label{fig.overview}Structure overview of the proposed code generator. Files coming from users are marked as yellow. Modules are marked as blue.}
\end{figure}

Our code generator employs a modular design, where each module functions similarly to the concept of a Pass in LLVM \cite{lattner2004llvm}. Multiple modules are sequentially stacked to form a pipeline that performs specific analyses and optimizations on each operator as the model flows through. Fig.~\ref{fig.overview} presents a structural overview of our code generator. The first two modules are dedicated to model optimization, while the DSA optimization module incorporates backend hardware specification to invoke DSA for specific operator acceleration. If no DSA is attached, this step can be skipped. The final generation module uses prepared templates to generate the main source code and saves model weights and activations in various formats.

\subsection{Modules}
%

The frontend module initially performs a forward pass on an input sample, such as an image, through the input model to cache intermediate activations for all hidden layers. Subsequently, the frontend module parses the input model file (currently supporting TFLite, with ONNX support under development) and extracts model information layer by layer, including weights and activations, which are converted into intermediate representations (IRs) stored in dictionary format. As the initial release, the current frontend supports key components of convolutional neural networks, e.g. \texttt{conv2d}, \texttt{dense}, \texttt{flatten}, and \texttt{maxpool}. Users are free to extend the parser to other operators.
For quantized models, quantization parameters like scaling factors and zero points are also extracted. Additionally, operator attributes, such as convolutional kernel size and padding, are stored within the weight attributes for potential DSA configuration. An example of an IR extracted by the frontend module is displayed in Table~\ref{tab.IR1}.
\begin{table}[htbp]
\centering
\caption{Layerwise IR extracted from the input model}\label{tab.IR1}
\begin{tabular}{lcccc}
\hline
\multicolumn{5}{c}{Layer Name}                                                         \\ \hline
\multicolumn{1}{c|}{Key}        & tensor  & scale & zero\_p & attribute                \\ \hline
\multicolumn{1}{l|}{i (input)}  & ndarray & FP32  & FP32    &                          \\
\multicolumn{1}{l|}{w (weight)} & ndarray & FP32  & FP32    & \multicolumn{1}{l}{dict} \\
\multicolumn{1}{l|}{b (bias)}   & ndarray & FP32  & FP32    &                          \\
\multicolumn{1}{l|}{o (output)} & ndarray & FP32  & FP32    & \multicolumn{1}{l}{}     \\ \hline
\end{tabular}
\end{table}

We observe that quantized models require scaling factors (often represented as FP32 values) to adapt quantized values to the full precision range in order to maintain accuracy. This not only necessitates floating-point units (FPU) in the backend but also introduces additional memory write overhead for rescaling computations between two accelerated layers, leading to increased inference time. Furthermore, many AI accelerators focus primarily on optimizing low-level operations, such as General Matrix Multiplications (GEMM) and convolutions, while delegating corresponding bias addition to the CPU for execution.

To address these issues, we propose two ahead-of-time (AOT) optimizations, referred to as weight tuning and bias fusion, and integrate them into the IR optimization module. These optimizations eliminate CPU-based rescaling and bias addition, thereby improving the efficiency of quantized model inference.
In principle, weight tuning can be applied to any quantized layer based on weighted sum operations, while bias fusion is applicable to layers with an enabled bias term. We detail this process targeting fully connected and 2D convolutional layers, whose single output is simply formulated as
\begin{equation}\label{eq0}
\sum x w + b = y,
\end{equation}
where $w$ and $b$ denote weights and bias, respectively.
The relationship between pre- and post-quantization values in input or output is stated as 
\begin{equation}\label{eq1}
d = s ( q_d -z) ,
\end{equation}
where $d$ indicates pre-quantized value, $q_d$ denotes post-quantized value, $s$ is the scaling factor, $z$ is the zero point.
In case that symmetric quantization with $z=0$ applied, we have 
\begin{equation}\label{eq2}
d = s \cdot q_d ,
\end{equation}
then the quantized output from equation~\eqref{eq0} can be represented by 
\begin{equation}\label{eq3}
\begin{aligned} 
\sum q_x q_w  \cdot \frac{s_x s_w}{s_y}+q_b \cdot \frac{s_x s_w}{s_y} &=q_y \\ 
(\sum q_x q_w+q_b)  \cdot \frac{s_x s_w}{s_y} &=q_y. 
\end{aligned}
\end{equation}
If \( q_y \) is the input to the following layer, a post-rescaling of the layer’s quantized output $(\sum q_x q_w+q_b)$ by the factor $ s_x  s_w / s_y $ is necessary, typically increasing additional CPU overhead. This overhead can be minimized by weight tuning, where the $ s_x  s_w / s_y $ factor is reformulated to a tuned value with a right shift of N bits, as shown in 
\begin{equation}\label{eq4}
\frac{s_x s_w}{s_y}=s_{tune} \gg N, s_{tune} \in [0.5, 1). 
\end{equation} 
Bit-shifting is computationally cheap from hardware side. 
Some DSA~\cite{hoppner2021spinnaker} even directly supports right-shifted outputs. Meanwhile, \( s_{tune} \) can be merged with \( q_w \) and \( q_b \), resulting in tuned weights and biases (\( q_w' = s_{tune} \cdot q_w \), \( q_b' = s_{tune} \cdot q_b \)), transforming \eqref{eq3} into 
\begin{equation}\label{eq5}
(\sum q_x q_w'+q_b') \gg N=q_y. 
\end{equation} 


\begin{figure}[htbp]
\begin{subfigure}[t]{0.48\textwidth}
\centering
\includegraphics[width=\textwidth]{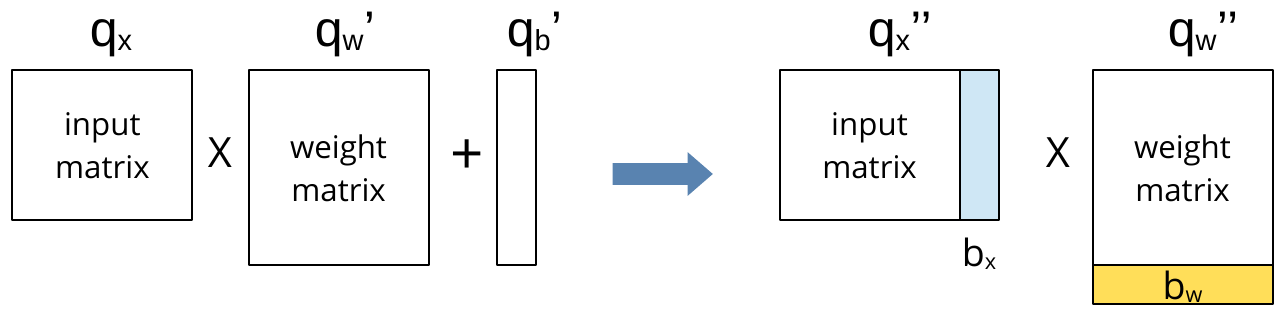} 
\caption{Fusion in fully-connected layer}\label{fig.bias_fusion.a}
\end{subfigure}
\begin{subfigure}[t]{0.48\textwidth}
\centering
\includegraphics[width=\textwidth]{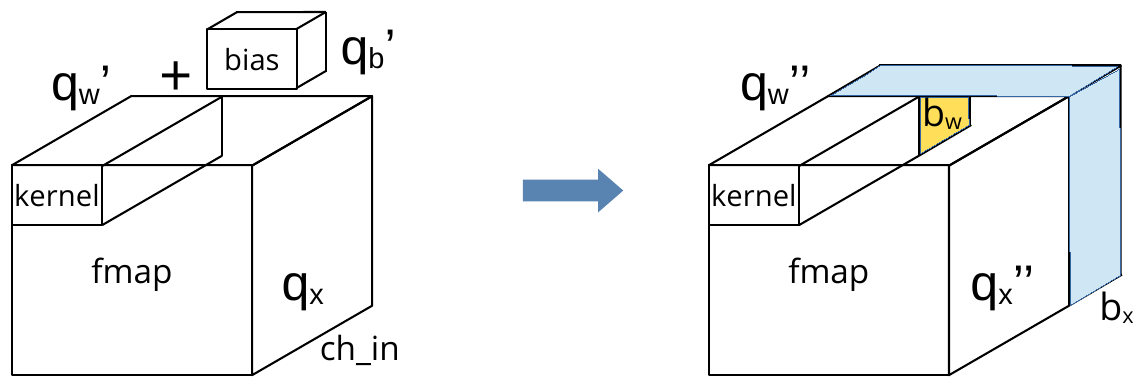} 
\caption{Fusion in convolutional layer}\label{fig.bias_fusion.b}
\end{subfigure}
\caption{Illustration of bias fusion}\label{fig.bias_fusion}
\end{figure}
Bias fusion further optimizes \eqref{eq5} by embedding bias addition into the weighted sum, as illustrated in Fig.~\ref{fig.bias_fusion}. For fully connected layers, this involves filling the input matrix with additional columns of a constant \( b_x \) and extending the weight matrix with additional rows of \( b_w \), where $q_b'=b_x b_w$. For convolutional layers, the input feature map is padded with additional channels of \( b_x \), and the convolutional kernels are expanded with additional channels of \( b_w \). This optimization further yields 
\begin{equation}\label{eq6}
\sum q_x'' q_w'' \gg N=q_y. 
\end{equation} 

Thus, peripheral calculations such as bias addition and rescaling are merged into DSA-driven operators, requiring only constant fills for certain extended input buffers. Although weight tuning and bias fusion introduce some truncation errors during optimization, experiments show that the impact on model accuracy remains minimal.

The DSA optimization module adapts the IRs for the target DSA backend, with its operations highly dependent on the specific registered DSA. Initially, the module cross-references the IR details with backend specifications to verify compatibility, checking whether any operators exceed the hardware’s support limits (e.g., feature map size constraints). Next, it conducts a dimension order check, as TFLite stores tensors in NHWC format (batch, height, width, channel), which may require conversion to NCHW format for certain DSAs. If the DSA requires a particular dimension ordering for weight storage, the weights are permuted accordingly at this stage as well. Memory alignment is subsequently verified. To maximize throughput, DSAs often fetch multiple values simultaneously, so width dimensions may require zero-padding to align with specific boundaries (e.g., 128-bit alignment). Finally, we employ a ping-pong buffer strategy to minimize allocated buffer memory, with a detailed example in section~\ref{sec.radarnet}.

Once the IR preparation is complete, C source code generation begins. Users are required to prepare C function templates specific to the backend, including the main function and layer functions for various supported layer types. 
For instance, Fig.~\ref{fig.template} depicts the code template of softmax layer for distinguishing the final class in a classifier.
The code generator sequentially parses IRs and translates them into corresponding function calls that populate the main function. Formulated tensors, such as \( q_x'' \) and \( q_w'' \) in Equation~\eqref{eq6}, are archived in multiple formats within the output directory, including C headers, JSON, and binary memory formats. Users can specify the desired output formats via the output configuration settings.

\begin{figure}[htbp]
\centering
\begin{lstlisting}
void layer_{{layer_i}}({{IR[layer_i]["i"]["tensor"].dtype}}_t* d_in, int cls_num, uint8_t* cls_out){
    //softmax is simplified by hardmax to select the class id as it is monotonically increasing.
    {{IR[layer_i]["i"]["tensor"].dtype}}_t output_i = *d_in;
    *cls_out = 0;
    for(uint8_t i = 1; i < cls_num; i++){
        if(d_in[i] > output_i){
            *cls_out = i;
            output_i = d_in[i];
        }   
    }   
}
\end{lstlisting}
\caption{\label{fig.template}Code template for Softmax layer using jinja2 library}
\end{figure}

\section{Evaluation}
To demonstrate how this code generator integrates seamlessly into a typical AI development workflow, we present a case study focused on a CNN-based radar target classification task. We detail the agile, codegen-based deployment on a custom radar DSP equipped with a machine learning accelerator capable of accelerating GEMMs and convolution operations. Our goal is to lower the barrier for future replication of this process and to encourage the development of new workflows for other backends.
\subsection{Case Study}\label{sec.radarnet}

\begin{figure}[htbp]
\begin{subfigure}[t]{0.48\textwidth}
\centering
\includegraphics[width=\textwidth]{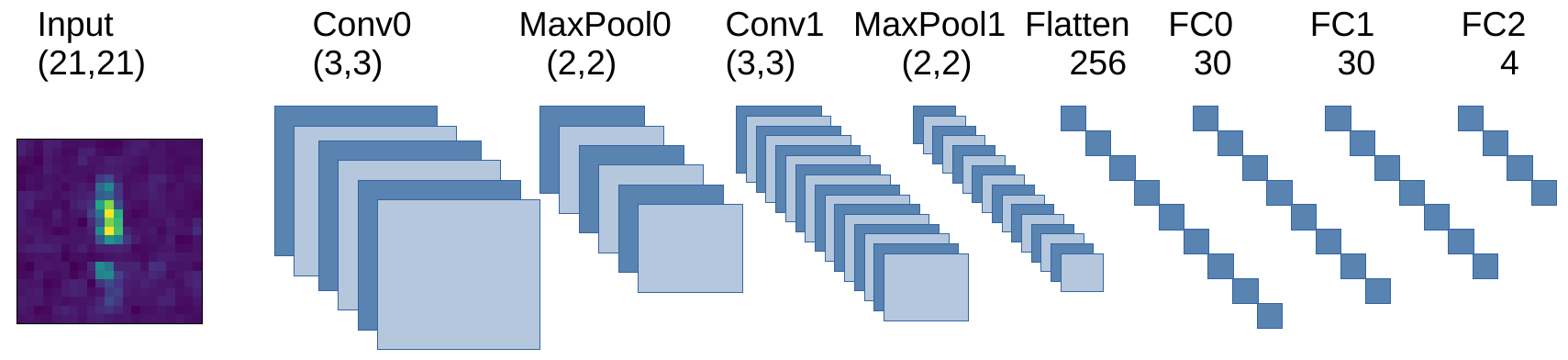} 
\caption{Structure overview}\label{fig.radarnet.a}
\end{subfigure}
\begin{subfigure}[t]{0.48\textwidth}
\centering
\includegraphics[width=\textwidth]{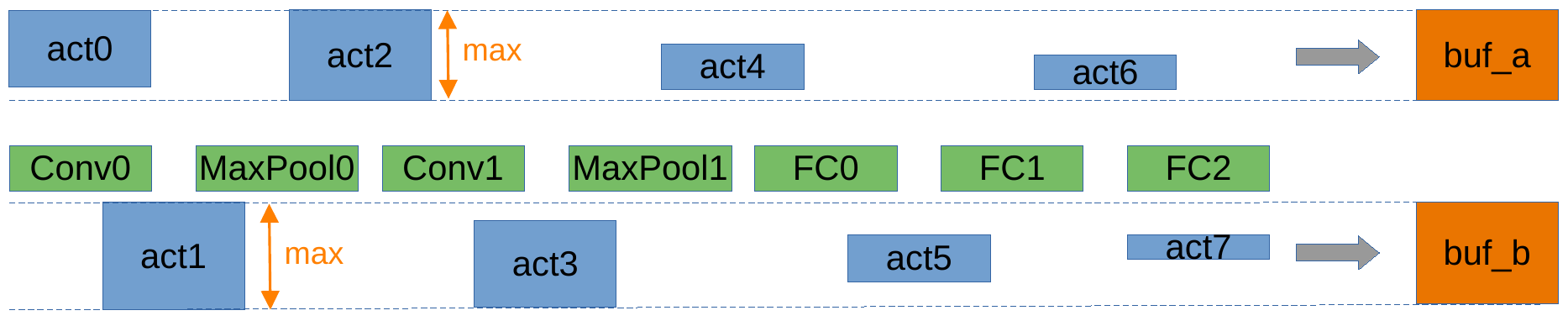} 
\caption{Ping-pong buffer strategy}\label{fig.radarnet.b}
\end{subfigure}
\caption{Illustration of RadarNet}
\end{figure}

To classify objects in millimeter-wave radar images, we developed a compact CNN model named as \textbf{RadarNet} with only 9,704 parameters that achieves 92\% accuracy on a four-class task for car, cyclist, pedestrian, and noise. The model consists of convolutional layers, max-pooling layers, and fully connected layers, as shown in Fig.~\ref{fig.radarnet.a}. Prior to the code generation, the model is statically quantized to int8 using TFLite. To minimize memory usage for intermediate activations, we implemented a ping-pong buffer strategy, which alternates storage of each layer’s output. Our code generator automatically calculates the buffer size based on the maximum tensor size collected for each buffer and reserves the corresponding memory regions in the generated code. For example, in Fig.~\ref{fig.radarnet.b}, \texttt{buf\_a} is sized equal to \texttt{act2}, while \texttt{buf\_b} fits \texttt{act1} exactly.



The employed backend device is a custom low-power radar DSP~\cite{gonzalez202312} that includes 25 processing elements (PEs): a peripheral core and 24 signal processing cores. Each processing core encompasses an embedded RISC-V core, a Machine Learning Accelerator (MLA), and 128 kB of memory. The MLA integrates 64 multiply-accumulate (MAC) units, which are leveraged to accelerate integer-based GEMM and convolution operations. This MLA also supports bit-shifted outputs, allowing direct computation of \( q_y \) in Equation~\eqref{eq6}.
\begin{figure}[htbp]
\centering
\includegraphics[width=0.9\linewidth]{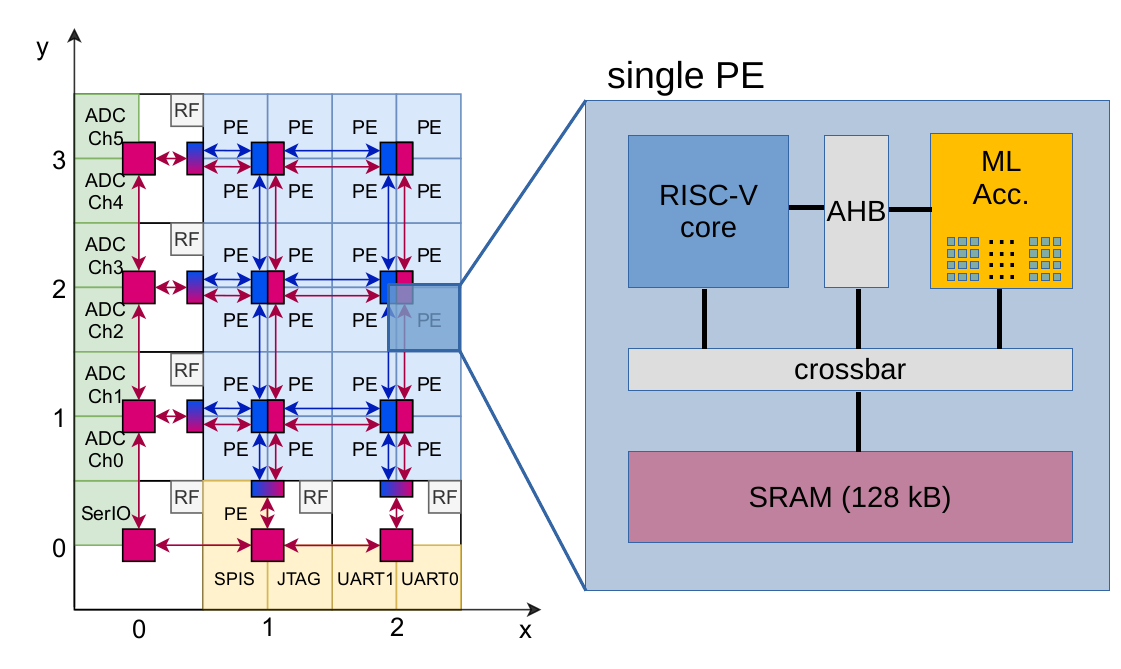}
\caption{\label{fig.radardsp}Overview of the adopted radar DSP as backend}
\end{figure}

In principle, once users prepare the required input files (as shown by yellow blocks in Fig.~\ref{fig.overview}), they can generate the model’s C source code with a single call of our code generator and then compile the code using the backend’s default toolchain.

\subsection{Result}
We compared three different code generation methods by deploying RadarNet on the aforementioned radar DSP, assessing hardware resources that do not pose bottlenecks for this model. The microTVM flow~\cite{liu2023deploying} was used as the baseline, as it relies solely on the RISC-V core for all computations without targeting any accelerator backend. In contrast, quantized convolutions (\texttt{qnn.conv}) were incorporated into the TVM flow through the Universal Modular Accelerator (UMA) interface~\cite{ahmadifarsani2023towards}. We extended it to include quantized dense layers (\texttt{qnn.dense}) as the comparison object, enabling full acceleration of both the convolutional and fully connected layers within RadarNet. These two TVM-based counterparts adopt static memory planning for tensor management.

\begin{table}[htbp]
\centering
\caption{Toolkit benchmark}\label{tab.benchmark}
\begin{tabular}{cccc}
\hline
            & \begin{tabular}[c]{@{}c@{}}exec. time\\ {[}us{]}\end{tabular} & \begin{tabular}[c]{@{}c@{}}binary size\\ {[}kB{]}\end{tabular} & \begin{tabular}[c]{@{}c@{}}manual\\ intervention\end{tabular} \\ \hline
microTVM~\cite{liu2023deploying}                & 4571  & 39.3  & -   \\
microTVM+UMA                                    & 1470  & 31.2  & 278+81   \\
\textbf{Our CodeGen}    & 630   & 25.0  & 677+211   \\ \hline
\end{tabular}
\end{table}

The benchmarking results are presented in Table~\ref{tab.benchmark}, showing single-inference time, compiled binary size, and the degree of manual intervention required for code generation. Compared to the baseline microTVM flow, integrating the MLA via the UMA interface reduces execution time and binary size to 32.2\% and 79.4\% of the original, respectively, while for our code generator, these figures are further reduced to 13.8\% and 63.6\%.

Comparing the last two rows, although both methods leverage the accelerator, our generator achieves more than double the speed of the \texttt{microTVM+UMA} approach. This performance gain is primarily due to two factors: first, while \texttt{microTVM+UMA} performs runtime dimension reordering (from NHWC to NCHW), our generator handles this conversion at compile time; second, whereas \texttt{microTVM+UMA} relies on the RISC-V core for rescaling after invoking the MLA, our generator completes this rescaling through compile-time weight tuning. Additionally, performing as many IR optimizations as possible at compile time minimizes binary size, enabling larger networks to fit on hardware with limited storage.

In this use case, the code lines in frontend modules and backend templates represent amount of manual intervention. Since our code generator implements all required operators in RadarNet and applies additional IR optimizations, while UMA focuses solely on operators mapped to accelerators, our overall code size is more than twice as large. However, the required incremental code size to register new operators is comparable to that of the TVM approach.

\subsection{Discussion}
As the initial release of the code generator, we have implemented key operators commonly used in CNNs, including \texttt{conv2d}, \texttt{dense}, \texttt{flatten}, and \texttt{maxpool}, within the frontend module. To support models with additional operators, users should extend the Python API in the frontend module, enabling the recognition and extraction of new operators into IRs. For deploying models on new backends, users need to prepare a set of \texttt{Jinja2} templates that define all operators used in the model and specify the unique characteristics of the new backend in the backend specification. Each individual set of templates is designed to fully exploit the specific hardware capabilities while ensuring isolation between different hardware backends. However, developing a comprehensive set of operator templates for a new backend requires a one-time effort.

From a methodological perspective, the compilation workflow of this code generator closely resembles that of TVM. Importantly, IR optimization techniques such as weight tuning and bias fusion, as employed in this work, are also applicable to the TVM compilation process. Nevertheless, the proposed code generator aims to provide a turnkey solution with minimal external dependencies and a lightweight executable runtime. In this regard, it takes an additional step beyond existing solutions.



\section{Conclusion}
In this work, we present a lightweight code generator that extracts IRs from a TFLite model and translates them into template-based C source code compatible with a backend chip. We demonstrate the complete workflow through a practical case study of radar target classification on a radar DSP in a real-world setting. Benchmarking against other state-of-the-art methods reveals that our generated code achieves the smallest executable size and lowest execution time.
To adapt it to another backend, the only effort is to prepare a set of corresponding templates for required operators.
We hope this approach can play a critical role in accelerating the deployment of machine learning models for heterogeneous backends. Moreover, as multi-core heterogeneous systems advance, we envision future research to focus on partitioning single operators into multiple tiles and mapping end-to-end workloads across multiple coupled cores and accelerators.